\documentclass[aps,twocolumn,pre]{revtex4-2}
\usepackage{amsthm}
\usepackage{ulem}
\usepackage{graphicx}
\usepackage{amsfonts}
\usepackage{amsmath}
\usepackage{bbold}
\usepackage{bm}
\usepackage{enumerate}
\usepackage{color}
\usepackage{comment}

\begin{document}
\title{Macroscopic dynamics of quadratic integrate-and-fire neurons subject to correlated noise}
\author{Hui Wang}
\affiliation{School of Physics and Astronomy, Yunnan University, Kunming, 650091, China}
\author{Chunming Zheng}
\email{cmzheng@ynu.edu.cn}
\affiliation{School of Physics and Astronomy, Yunnan University, Kunming, 650091, China}

\begin{abstract}
The presence of correlated noise, arising from a mixture of independent fluctuations and a common noisy input shared across the neural population, is a ubiquitous feature of neural circuits, yet its impact on collective network dynamics remains poorly understood. We analyze a network of quadratic integrate-and-fire neurons driven by Gaussian noise with a tunable degree of correlation. Using the cumulant expansion method, we derive a reduced set of effective mean-field equations that accurately describe the evolution of the population's mean firing rate and membrane potential. Our analysis reveals a counterintuitive phenomenon: increasing the noise correlation strength suppresses the mean network activity, an effect we term \textit{correlated-noise-inhibited spiking}. Furthermore, within a specific parameter regime, the network exhibits metastability, manifesting itself as spontaneous, noise-driven transitions between distinct high- and low-activity states. These results provide a theoretical framework for reducing the dynamics of complex stochastic networks and demonstrate how correlated noise can fundamentally regulate macroscopic neural activity, with implications for understanding state transitions in biological systems.
\end{abstract}

\maketitle

\section{Introduction}
The constructive role of correlated fluctuations, far from being a simple nuisance, is a fundamental driver of collective behavior across diverse physical and biological systems. This phenomenon is exemplified by the Moran effect in theoretical ecology, where environmentally correlated noise can synchronize populations \cite{moran1953statistical,grenfell1998noise}, and in systems of coupled oscillators, where it can enhance  synchronization \cite{pikovsky1984Synchronization,zhou2002noise, teramae2004robustness, goldobin2005synchronization,nakao2007noise,gong2019repulsively}. Similarly, in active matter, spatially correlated noise can induce phase transitions and robust collective motion \cite{zheng2022noise}. A unifying principle across these domains is that a common noise component, shared across an ensemble, can serve as an effective common input at the macroscopic level, inducing correlated or entrained dynamics qualitatively different from those produced by independent fluctuations.

This principle is profoundly relevant in neuroscience \cite{lampl1999synchronous,doiron2016mechanics}. Neuronal circuits operate in a regime of intense synaptic background activity, which is a primary source of intrinsic noise. A critical feature of this synaptic noise is its correlation structure: while some fluctuations are independent across neurons, a significant portion is shared, arising from common presynaptic inputs or neuromodulation. Numerous experimental and theoretical studies have shown that shared or correlated noise can strongly influence neuronal variability and collective behavior.
In sensory systems, common fluctuations can enhance or degrade population coding depending on the stimulus \cite{averbeck2006neural,moreno2014information}. At the network level, correlated noise can modulate firing-rate variability and pairwise correlations \cite{renart2010asynchronous}, affect the power spectrum of population activity \cite{de2007correlation}, and can also change the reliability and gain of single-neuron responses \cite{salinas2000impact}, and shape information transmission across neural ensembles.
Consequently, understanding how correlated noise shapes network dynamics is not merely a theoretical exercise but is essential for deciphering the mechanisms underlying neural coding, plasticity, and decision-making.

The analytical challenge of modeling noise correlations is formidable. While the statistical effects of correlated noise are tractable for networks of uncoupled leaky integrate-and-fire neurons \cite{lindner2005theory,tchumatchenko2010correlations}, the interplay between synaptic coupling and correlated noise in shaping dynamics remains poorly understood. A major breakthrough for analyzing coupled oscillators was the development of the Ott-Antonsen (OA) ansatz \cite{ott2008low}, which provides an exact reduction for the macroscopic dynamics of large ensembles. The quadratic integrate-and-fire (QIF) model is a canonical representation of class I neural excitability \cite{izhikevich2007dynamical,ermentrout1986parabolic}, which has been broadly studied theoretically and experimentally for excitable systems \cite{byrne2020next,rabuffo2021neuronal,di2018transition,pietras2023exact,goldobin2024discrete,ness2025validity,gerster2021patient,laing2015exact,jia2023bioelectrical}.
For coupled QIF neurons, the Lorentzian ansatz formalism \cite{montbrio2015macroscopic} yields exact low-dimensional equations for the mean firing rate and mean membrane potential of the network. However, the OA ansatz is restricted to scenarios with fully correlated (common) noise or specific noise distributions (e.g., Cauchy). It breaks down under the more biologically prevalent case of independent or partially correlated Gaussian noise, severely limiting its applicability to realistic neural systems.

To overcome this limitation, the cumulant expansion method has emerged as a powerful alternative \cite{tyulkina2018dynamics,goldobin2021reduction}. This approach does not rely on the restrictive assumptions of the OA ansatz and provides a systematic framework for deriving approximate, yet highly accurate, low-dimensional descriptions of population dynamics driven by Gaussian noise.

In this paper, we bridge this gap by employing the cumulant expansion method to investigate the macroscopic dynamics of an excitatory coupled network of QIF neurons subjected to partially correlated Gaussian noise. We focus on how the degree of noise correlation modulates collective network states. Specifically, we uncover and characterize two key phenomena: (i) a novel suppression of population activity with increasing correlation strength, which we term correlated-noise-inhibited spiking, and (ii) robust metastability, featuring coherent noise-driven switching between high- and low-activity states. Our work establishes a theoretical framework for reducing the complex stochastic dynamics of coupled neural populations to a tractable system, offering novel insights into how correlated fluctuations at the synaptic level can dictate large-scale, functional dynamics in neural circuits.

\section{Model}
We consider a large-scale network of $N$ globally coupled quadratic integrate-and-fire (QIF) neurons, a canonical model for Class I neural excitability. The dynamics of the membrane potential $V_j$ for neuron $j$ ($j=1,\ldots,N$) are given by
\begin{equation}
\dot{V}_j = V_j^2 + I_j, \quad \text{with} \quad I_j = \eta_j + J s(t) + I_0 + \xi_j(t).
\label{Eq:Model_V}
\end{equation}
A spike is emitted when $V_j$ diverges through an upper threshold ($V_j \to +\infty$), upon which it is reset to $-\infty$. This infinite threshold-reset limit preserves mathematical tractability while capturing the essential excitable dynamics of the neuron. The total input current $I_j$ comprises four components: (i) a quenched heterogeneous parameter $\eta_j$, representing the intrinsic excitability of each neuron; (ii) the synaptic input from the network, with global coupling strength $J$; (iii) a constant external drive $I_0$; and (iv) a fluctuating current $\xi_j(t)$ representing correlated noise.

The heterogeneity in intrinsic excitabilities $\eta_j$ is modeled by drawing them from a Lorentzian (Cauchy) distribution,
\begin{equation}
g(\eta) = \frac{1}{\pi} \frac{\Delta}{(\eta - \bar{\eta})^2 + \Delta^2},
\end{equation}
with median $\bar{\eta}$ and half-width at half-maximum $\Delta$.

The synaptic activation $s(t)$ represents the population-averaged input from all incoming spikes. It is defined as the convolution of the spike train of the network with a normalized synaptic kernel $a_\tau(t)$:
\begin{equation}
s(t) = \frac{1}{N} \sum_{j=1}^{N} \sum_{k} \int_{-\infty}^{t} dt' a_\tau(t - t')\delta(t' - t_j^k),
\end{equation}
where $t_j^k$ denotes the $k$-th spike time of neuron $j$, $\delta(\cdot)$ is the Dirac delta function, and the kernel $a_\tau(t) = (1/\tau) e^{-t/\tau} \Theta(t)$ has a characteristic synaptic time constant $\tau$ ($\Theta(t)$ is the Heaviside function). To describe the collective network activity, we introduce the macroscopic (population-averaged) order parameters
\begin{equation}
r(t) = \frac{1}{N}\sum_{j=1}^{N}\sum_{k}\delta(t - t_j^k), \qquad
v(t) = \frac{1}{N}\sum_{j=1}^{N} V_j(t),
\end{equation}
which represent the instantaneous population firing rate and the mean membrane potential, respectively.
Throughout this paper, we consider the limit of fast synapses , $\tau \to 0$. In this limit, synaptic activation $s(t)$ immediately follows the mean firing rate, i.e. $s(t) = r(t)$.

A central focus of this study is on the role of correlated noise. The noise term $\xi_j(t)$ for each neuron is constructed from two independent Gaussian white noise sources:
\begin{equation}
\xi_j(t) = \sqrt{c}\xi_c(t) + \sqrt{1-c}\tilde{\xi}_j(t).
\label{Eq:noise_decomposition}
\end{equation}
Here, $\xi_c(t)$ is a common noise component shared by all neurons, and $\tilde{\xi}_j(t)$ is an independent noise component specific to each neuron. Both $\xi_c(t)$ and $\tilde{\xi}_j(t)$ have zero mean and correlation $\langle \xi_\alpha(t) \xi_\alpha(t') \rangle = 2D \delta(t-t')$ ($\alpha = c$ or $j$), ensuring that the total noise intensity for each neuron is $D$. The parameter $c \in [0,1]$ is the noise correlation coefficient, which continuously tunes the noise from being purely independent across neurons ($c=0$) to being perfectly common ($c=1$).

In the thermodynamic limit ($N \to \infty$), the state of the system is described by a probability density function $P(V,t,|\eta)$ for the membrane potential. For a given realization of the common noise $\xi_c(t)$, the evolution of this density is governed by a stochastic Fokker–Planck equation:
\begin{equation}
\frac{\partial P(V, t | \eta)}{\partial t}
= -\frac{\partial}{\partial V}\big[F(V) \, P(V, t | \eta)\big]
+ (1-c)D \frac{\partial^2 P(V, t | \eta)}{\partial V^2}.
\label{Eq:FKE_V}
\end{equation}
where the drift term is
\begin{equation}
F(V) = V^2+\eta + J r(t) + I_0 + \sqrt{c}\xi_c(t).
\end{equation}
The diffusion term in Eq.~\eqref{Eq:FKE_V} makes the OA ansatz problematic. To overcome this fundamental obstacle, we employ the cumulant expansion method \cite{tyulkina2018dynamics,goldobin2021reduction}. This approach provides a systematic framework for deriving approximate low-dimensional ordinary differential equations that govern the evolution of the macroscopic order parameters, i.e. the mean firing rate $r(t)$ and the mean membrane potential $v(t)$, effectively averaging over the independent noise and the heterogeneity. In the following section, we detail this derivation, which allows us to incorporate both quenched heterogeneity and partially correlated noise in a unified and tractable manner.

\section{Effects of correlated noise on macroscopic dynamics}
In this section, we derive the macroscopic dynamics of the neuronal network and examine how correlations in the external noise shape both the collective dynamics and the statistical properties of the system. Following the cumulant expansion method \cite{goldobin2021reduction}, we obtain an effective Langevin description of the macroscopic variables, i.e. the mean firing rate and the mean membrane potential, together with the corresponding Fokker–Planck equation.

We define the logarithm of the characteristic function $\Phi_\eta(k)=\ln\langle e^{ikV}\rangle$ and average over the intrinsic excitability, i.e. $\Phi(k)=\int\Phi_\eta g(\eta)d\eta$, to obtain the evolution equation
\begin{equation}
    \partial_t\Phi=ik[H-\partial^2_k\Phi-(\partial_k\Phi)^2]-|k|\Delta-(1-c)Dk^2.
    \label{Eq:Charac_Func}
\end{equation}
Here, $H=Jr(t)+I_0+\bar{\eta}+\sqrt{c}\xi_c(t)$.
By introducing the pseudo-cumulants $W_n=q_n+ip_n$, which are the coefficients of the small $|k|$ expansion of $\Phi$, one can truncate the hierarchy of equations of $W_n$ at second order by neglecting terms with $n \geq 3$ \cite{goldobin2021reduction}. We further employ a quasi-stationary closure by assuming that the second pseudo-cumulants $q_2$ and $p_2$ evolve on a faster timescale than the macroscopic variables $r$ and $v$, which allows us to set $\dot{q}_2=\dot{p}_2=0$ (see Appendix~\ref{Appendix} for details). This closure corresponds to a reduction to the slow manifold, where the higher-order moments are enslaved by the dynamics of the mean firing rate and membrane potential. As a result, we arrive at the following Langevin equations for the mean firing rate $r$ and the mean membrane potential $v$:
\begin{equation}
\begin{aligned}
\dot{r} &= (\Delta+p_2) /\pi + 2rv, \\
\dot{v} &= v^2 + \overline{\eta} + Jr + I_0 - \pi^2 r^2 + q_2 + \sqrt{c}\xi_c(t).
\end{aligned}
\label{Eq:Langevin_r_v}
\end{equation}

\begin{figure}
\centering
\includegraphics[width=0.49\textwidth]{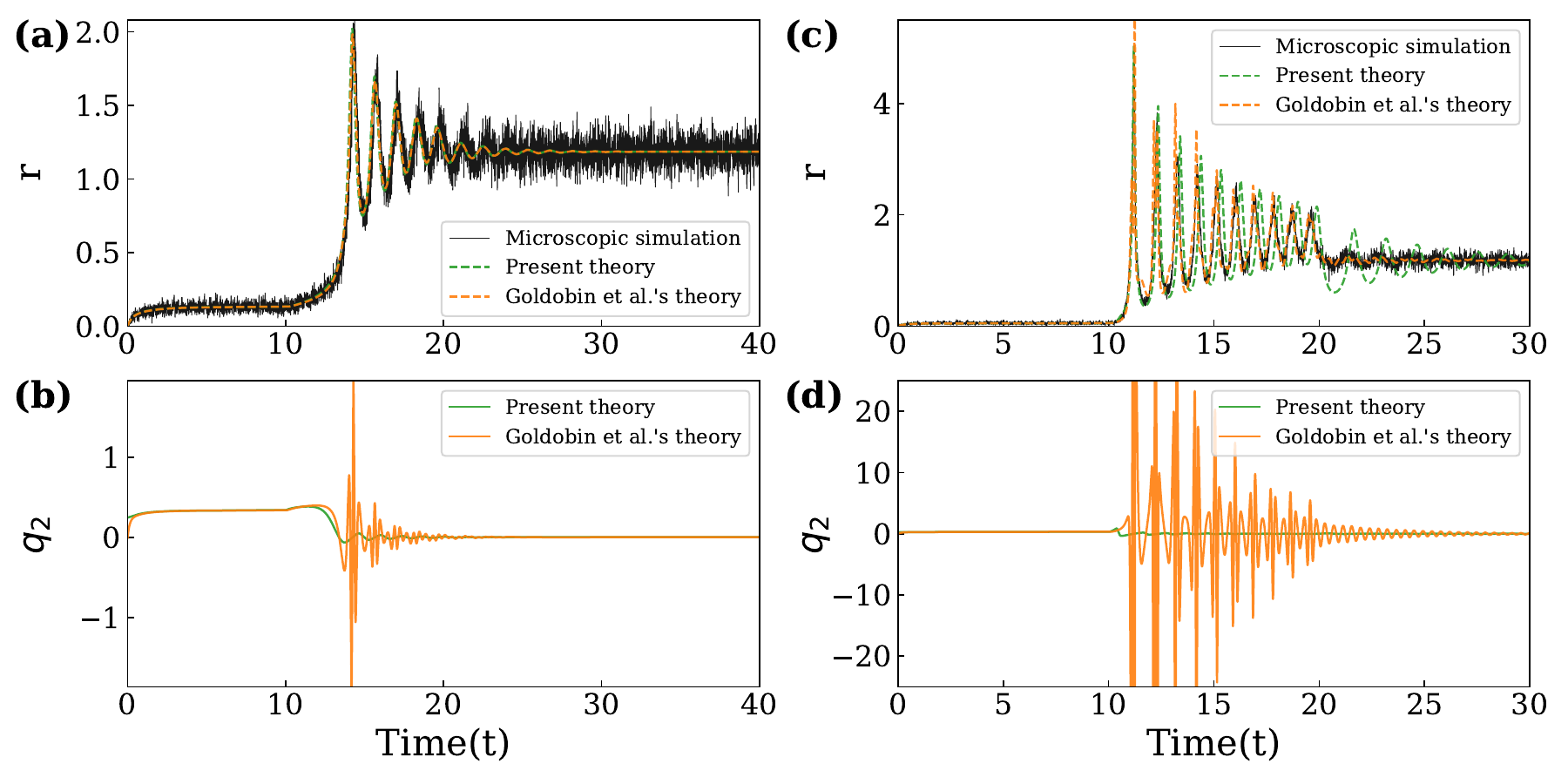}
\caption{Transient dynamics of an ensemble of $10^{4}$ coupled QIF neurons driven solely by independent noise ($c=0$). (a) and (b) depict the population-averaged firing rate $r(t)$ and the second cumulant $q_2(t)$ for $\Delta=1$, respectively. External step current input is set as $I_0=0.5$ for $t\in[0,20]$ and $I_0=0$ otherwise. (c) and (d) are the population-averaged firing rate $r(t)$ and the second cumulant $q_2(t)$ for $\Delta=0.5$, respectively. External step current input in this case is set as $I_0=5$ for $t\in[0,20]$ and $I_0=0$ otherwise. The solid black curves represent microscopic simulations, while the green and orange curves correspond to the present truncation ansatz Eq.~\eqref{Eq:Langevin_r_v} and the truncation proposed by Goldobin et al. \cite{goldobin2021reduction}, respectively. Other common system parameters used in (a)-(d): coupling strength $J=15$ and Lorentzian distribution parameters $\bar{\eta}=-3.94$. The integration time step is $dt=10^{-4}$.}
\label{Fig:Trunc_ansatz}
\end{figure}

Here, $p_2 = \frac{\pi r(1-c)D}{2(v^2 + \pi^2r^2)}$, $q_2 = -\frac{ v(1-c)D}{2(v^2 + \pi^2r^2)}$. Note that Eq.~\eqref{Eq:Langevin_r_v} reduces to the Lorentzian ansatz \cite{montbrio2015macroscopic} for $c=1$, that is, in the absence of uncorrelated noise component. To verify that the truncation ansatz in Eq.~\eqref{Eq:Langevin_r_v} faithfully reproduces the low-dimensional dynamics of the neuronal ensemble, we compare the transient responses obtained from microscopic simulations with those predicted by two truncation schemes. For illustration, we consider the case of purely independent noise ($c=0$); the same approach applies for arbitrary $c \in [0,1]$. The proposed truncation ansatz constitutes a further simplification of the cumulant expansion approach introduced by Goldobin et al. \cite{goldobin2021reduction}, while retaining the essential statistical moments governing the population dynamics. As shown in Fig.~\ref{Fig:Trunc_ansatz}(a), the mean firing rate $r(t)$ predicted by our reduced description is in excellent quantitative agreement with microscopic simulations. For clarity, we focus the quantitative comparison on $r(t)$ and the second cumulant $q_2(t)$; the mean voltage $v(t)$ and $p_2(t)$ show qualitatively similar behavior and are omitted. Figure~\ref{Fig:Trunc_ansatz}(b) demonstrates that $q_2(t)$ relaxes to its stationary state more rapidly than $r(t)$ following a transient input, thereby validating the quasi-stationary assumption ($\dot{q}_2 = \dot{p}_2 = 0$) that underpins our closure. However, it is important to note the limitation of our approach: its quantitative accuracy for transient dynamics diminishes in parameter regimes where  noise becomes dominant over heterogeneity. This is demonstrated in Fig.~\ref{Fig:Trunc_ansatz}(c, d), where for a reduced heterogeneity ($\Delta=0.5$), the transient dynamics of both $r(t)$ and $q_2(t)$ show a discernible discrepancy compared to the more complete 4D description of Goldobin et al. \cite{goldobin2021reduction}. Despite this, our 2D truncation ansatz still reliably captures the long-time stationary behavior of the macroscopic dynamics.

The corresponding Fokker–Planck equation governing the probability density $\rho(r,v,t)$ of the macroscopic variables takes the form
\begin{equation}
\begin{aligned}
\partial_t \rho(r,v,t) 
= & - \frac{\partial}{\partial v}\!\Big[\big(v^2 + \overline\eta + J r + I_0 - \pi^2 r^2 + q_2\big)\,\rho\Big] \\
  & - \frac{\partial}{\partial r}\!\Big[\big((\Delta+p_2)/\pi + 2 r v \big)\,\rho\Big] \\
  & + cD \,\frac{\partial^2 \rho}{\partial v^2}.
\end{aligned}
\label{Eq:FKE_corr_noise}
\end{equation}
This formulation establishes a direct link between microscopic noise correlations and the emergent macroscopic dynamics of the neuronal network.

In the limiting case of purely common noise ($c=1$), the Langevin and Fokker–Planck equations reduce exactly to the Lorentzian ansatz, with the contributions from $p_2$ and $q_2$ vanishing. However, in the general case, obtaining an analytical solution of the two-dimensional Fokker–Planck equation \eqref{Eq:FKE_corr_noise} is intractable. We therefore compute the probability density function $\rho(r,v)$ numerically using a finite-difference scheme. The time evolution is discretized with a second-order Runge-Kutta method (RK2) using a time step of $\Delta t = 10^{-4}$, and Dirichlet boundary conditions are applied. The computational domain for the variables is set to $r \in [0, 4]$ with 200 grid points and $v \in [-5, 5]$ with 250 grid points. To ensure a valid probability distribution, the solution is normalized at each time step by dividing by the numerical integral $\iint \rho(r,v)drdv$.
\begin{figure}
\centering
\includegraphics[width=0.5\textwidth]{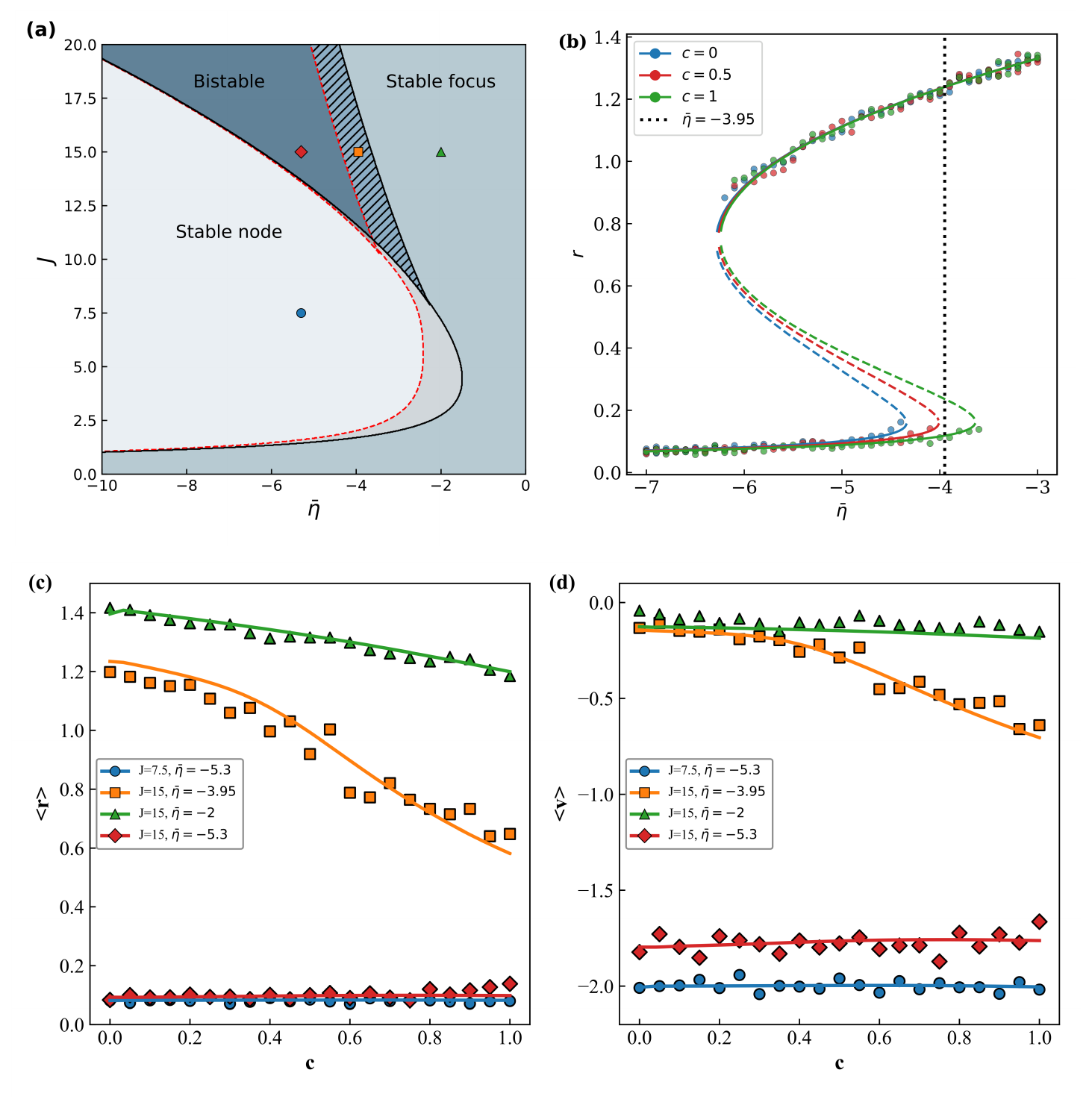}
\caption{(a) Phase diagrams for uncorrelated ($c=0$; red dashed boundaries) and fully correlated ($c=1$; black solid boundaries) noise. Dynamical regimes are color-coded. (b) Bifurcation diagram of the mean firing rate $r$ versus the median excitability $\bar{\eta}$. Theory (lines) agrees with microscopic QIF neuron simulations (dots). Both are obtained from the deterministic part of Eq.~\eqref{Eq:Langevin_r_v}, where the explicit noise term $\sqrt{c}\xi_c(t)$ is omitted, while the coefficients $q_2$ and $p_2$ retain their dependence on the correlation
coefficient $c$ and noise intensity $D$. (c, d) Dependence of the (c) mean firing rate and (d) mean membrane potential on the correlation strength $c$. Theory (solid curves) and simulation (symbols) are shown for parameters selected in (a) (matched by color/shape). Fixed parameters: $\Delta = 1$, $I_0 = 0.5$, $D=1$ and $dt=10^{-4}$ for simulations.}
\label{Fig:Diagram_J_eta}
\end{figure}

The impact of noise correlation on the dynamics of the system is summarized in Fig.~\ref{Fig:Diagram_J_eta}. Figure \ref{Fig:Diagram_J_eta}(a) shows phase diagrams in the $(\bar{\eta}, J)$ parameter space for the two limiting cases $c=0$ (uncorrelated noise, red dashed boundaries) and $c=1$ (fully correlated noise, black solid boundaries). A key observation is that the entire phase diagram shifts to the right as the correlation coefficient $c$ increases from $0$ to $1$, indicating that noise correlation plays a non-trivial role in shaping the collective dynamics.

To examine this effect across different dynamical regimes, we focus on a specific transition region where the influence of $c$ is most pronounced. This region, indicated by the hatched area in Fig.~\ref{Fig:Diagram_J_eta}(a), exhibits the largest shift between the $c=0$ and $c=1$ phase boundaries. We select a representative point within this region, $(\bar{\eta},J)=(-3.95, 15)$ (orange square), for detailed study. As shown in the bifurcation diagram of the mean firing rate $r$ of the network versus $\bar{\eta}$ in Fig.~\ref{Fig:Diagram_J_eta}(b), a stable focus (high-activity state) for $c=0$ enters a bistable regime as $c$ increases toward 1. For comparison, we choose parameter values in different dynamical regimes and plot the mean firing rate and mean membrane potential as a function of the correlation coefficient $c$, respectively; see Fig.~\ref{Fig:Diagram_J_eta} (c) and (d), both by simulations of the QIF network and by numerical calculation via the Fokker-Planck equation. The mean values are defined as $\langle r\rangle=\int\int r\rho(r,v)drdv$ and $\langle v\rangle=\int\int v\rho(r,v)drdv$, with $\rho(r,v)$ normalized density, in the stationary state. We can observe that the regime with the focus-to-bistability transition as varying the coefficient $c$ has the largest decrease in the mean activity for the QIF network.
We will analyze the mechanisms underlying this bistability transition in detail in the next section.

\section{Collective Metastability Driven by Correlated Noise}
\begin{figure*}
\centering
\includegraphics[width=0.95\textwidth]{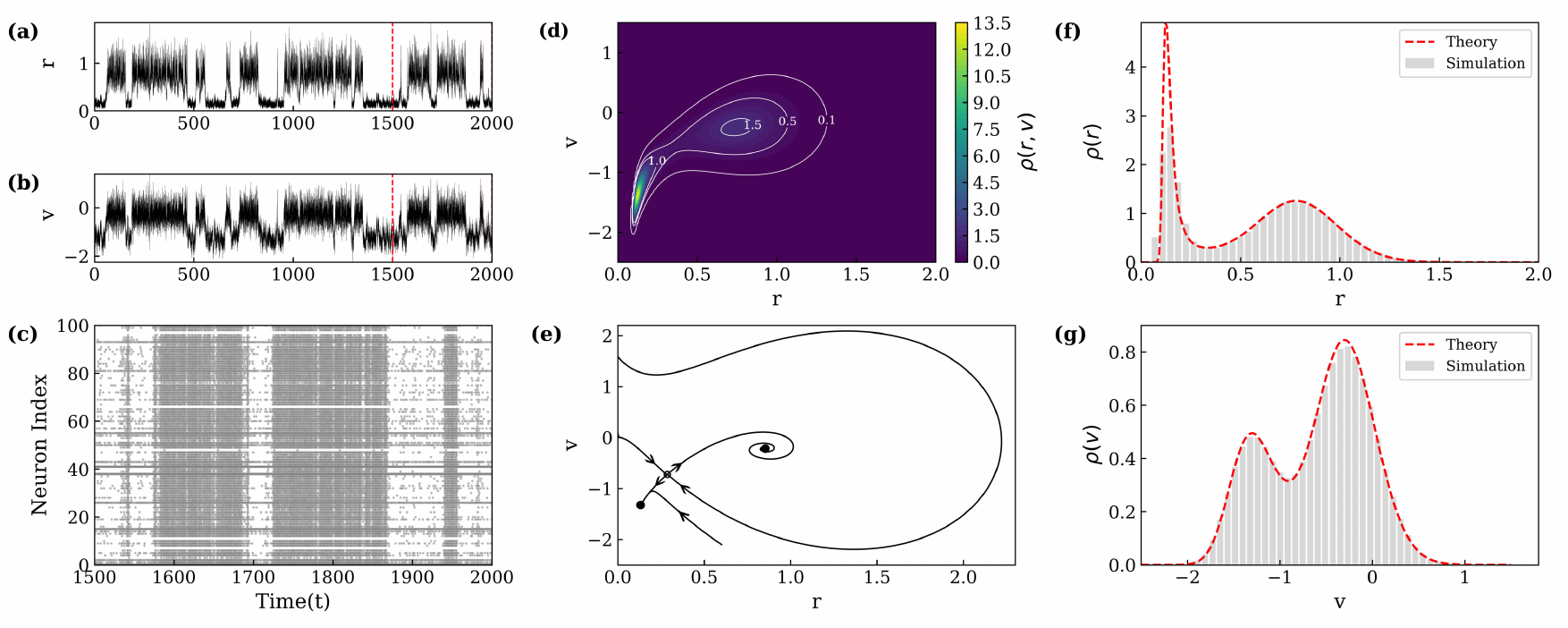}
\caption{This figure illustrates the changes in (a) firing rate and (b) mean membrane potential over time, as well as (c) the raster plots for the time interval $t\in[1500, 2000]$. The red dashed vertical line at $t=1500$ in panels (a) and (b) indicates the beginning of this interval. (d) Stationary probability density $\rho(r,v)$ obtained numerically from the two-dimensional Fokker–Planck equation Eq.~\eqref{Eq:FKE_corr_noise}. The color scale indicates the density magnitude, and white contours highlight regions of equal probability. (e) Deterministic phase portrait of the corresponding system.  (f) and (g) give the stationary probability distribution of the mean firing rate and mean membrane potential of the whole neuron network, respectively. The histograms depict microscopic simulations involving $10^4$ neurons over $2000$ seconds, while the theoretical results are illustrated by the red dashed lines. Parameters are set as $\overline{\eta} = -4.04$, $\Delta = 1$, $J = 15$, $I_0 = 0.5$, $c = 0.1$ and $D=1$.}
\label{Fig:Metastability}
\end{figure*}
Metastability, i.e. bistable transitions due to noise or dynamical structure of the neuronal network, has attracted considerable attention in neuroscience \cite{brinkman2022metastable,rossi2025dynamical,jercog2017up,litwin2012slow,mazzucato2019expectation,taher2020exact}, both from theoretical and experimental perspectives. This phenomenon is thought to underpin crucial cognitive functions such as working memory, decision-making, and perceptual switching, where neural circuits must maintain information yet remain flexible to new inputs. In the QIF network considered here, one can also observe robust metastable dynamics characterized by recurrent, spontaneous transitions between high- and low-activity states. To demonstrate this phenomenon, we fix the correlation coefficient at $c=0.7$ and choose other parameters ($\overline{\eta}=-3.95$, $\Delta=1$, $J=15$, $I_{0}=0.5$) such that the system operates within the bistable regime, as depicted in Fig.~\ref{Fig:Metastability}.

In this regime, the macroscopic state of the network undergoes stochastic switching, driven by the common component of the noisy fluctuations. The dynamics can be conceptualized as a particle randomly wandering in a double-well potential, where the two minima correspond to a stable focus (associated with the high-activity state) and a stable fixed point (representing the low-activity state). The finite correlation ($c>0$) in the noise provides a collective drive that pushes the entire population coherently between these two attractors.

Figure~\ref{Fig:Metastability} provides a comprehensive illustration of this metastable behavior. Panels (a) and (b) reveal that the population-averaged firing rate $r(t)$ and membrane potential $v(t)$ undergo spontaneous, noise-driven transitions between a quiescent and an active state. The transitions in firing rate and voltage are highly correlated. Moreover, the duration of residence in each state is highly variable, indicative of an escape process governed by noise. This collective dynamical feature is visibly reflected in the population spiking activity of the raster plot in (c), which shows clear epochs of sustained, low-activity state alternating with periods of intense high-activity state. The sharp transitions in the raster plot confirm that these switches are coherent across the neuronal population, driven by the common noise source.

We also plot the stationary probability density $\rho(r,v)$ by numerical calculation of the Fokker-Planck equation \eqref{Eq:FKE_corr_noise}. As shown in panel (d), the probability density manifests a sharp peak around the stable node region and shows a second peak around the stable focus region. Note that the stable node and stable focus regions represent the corresponding two stable fixed points for the deterministic part of the system. We plot the phase portrait of the deterministic dynamics in panel (e), where the stable focus and the stable node are indicated by black filled circles, separated by the unstable saddle indicated by an open circle. This specific phase-space structure gives rise to the asymmetric stationary probability density $\rho(r,v)$ shown in panel (d). The density is sharply concentrated near small $r$ values, with a secondary region of elevated probability around the focus, leading to a pronounced bimodality in the marginal distribution $\rho(r)$ and $\rho(v)$ (panel (f) and (g)). The stationary probability distributions for the firing rate and membrane potential, computed from the numerical solution of the Fokker-Planck equation \eqref{Eq:FKE_corr_noise}, show remarkable agreement with the histograms generated from extensive microscopic simulations of the full spiking network. This agreement is non-trivial; the bimodal shape of the distributions directly quantifies the metastability, with the two peaks corresponding to the two preferred states between which the system fluctuates. This close match between theory and simulation provides strong confirmation that our reduced model accurately encapsulates the essential stochastic dynamics of the full network. 

We emphasize that when the common noise component becomes large, the marginal distributions of both $r$ and $v$ may appear unimodal even though the deterministic phase diagram indicates bistability. This unimodality reflects noise-induced smoothing of the macroscopic variables rather than the absence of underlying multiple fixed points. Furthermore, our reduced mean-field description, comprising the Langevin equations for $r$ and $v$ (Eqs.~\eqref{Eq:Langevin_r_v}) and the associated Fokker-Planck equation (Eq.~\eqref{Eq:FKE_corr_noise}), provides a theoretical framework for making concrete quantitative predictions about macroscopic fluctuations. For instance, one could calculate and compare their variances or correlation times in the high- and low-activity metastable states, a promising direction for future work.

\section{Conclusions}
The response of individual neurons to correlated noise has been a subject of extensive study; however, its collective impact on the macroscopic dynamics of large neural networks remains an open question. In this work, we have addressed this gap by employing a cumulant expansion approach to derive low-dimensional mean-field descriptions for a network of QIF neurons driven by correlated noise. Our analysis has uncovered and characterized two key emergent phenomena: correlated-noise-inhibited spiking and metastability.

We have systematically mapped the dynamical phases of the system, demonstrating that increasing the noise correlation coefficient $c$, effectively amplifying the common noise component, induces a transition from a monostable regime (characterized by a stable focus) to a bistable regime. This structural change in phase space underlies the counterintuitive phenomenon of correlated-noise-inhibited spiking, where the mean population firing rate decreases as the correlation strength increases. This finding reveals a novel mechanism for global activity regulation, in which the coherence of external fluctuations, rather than their intensity, can suppress network-wide spiking.

Furthermore, we have elucidated the distinct mechanistic roles of the noise components within the bistable regime. The individual noise dictates the topology of the underlying phase space, determining the existence and location of attractors (e.g., a high-firing-rate state and a low-firing-rate state). In contrast, the common noise acts as a macroscopic driving force, provoking spontaneous and coherent transitions between these attractors, thus giving rise to metastable dynamics. The quantitative agreement between our theoretical framework—specifically, the stationary solutions of the derived Fokker-Planck equation—and large-scale microscopic simulations validates the efficacy of the cumulant expansion method. 
Note that retaining more higher-order cumulants can, in principle, improve the accuracy of the reduction~\cite{goldobin2024macroscopic}. However, this requires introducing additional macroscopic equations, which substantially increases the computational cost, especially for coupled networks with clustered structures or correlated noise as considered in the present work. Hence, the present approach can be viewed as a simplified and computationally efficient extension of the cumulant-based reduction by Goldobin et al. \cite{goldobin2021reduction}, offering a practical framework for analyzing collective dynamics of large spiking networks, particularly in the regime of moderate and low noise intensity.

Although our study offers novel insights into the low-dimensional dynamics of complex neural fluctuations, several avenues remain for future exploration. The method's accuracy could be further tested against more complex network structures, including sparse connectivity and synaptic heterogeneity. Moreover, extending this framework to incorporate additional biological realism, such as short-term synaptic plasticity or different neuronal time constants, would be a natural and important next step. Ultimately, by bridging the gap between microscopic noise correlations and macroscopic network dynamics, this work provides a foundational step towards understanding how correlated activity on one scale manifests as complex, functional dynamics on another.

\section{Acknowledgements}
We are grateful to the anonymous referees for their helpful comments and insightful suggestions that improved this work. We acknowledge funding from the Yunnan Fundamental Research Projects (Grant No. 202401AU070216) and the Scientific Research Fund Project of the Education Department of Yunnan Province (Grant No. 2024J0012).

\appendix
\section{Derivation of the macroscopic equation \eqref{Eq:Langevin_r_v}}
\label{Appendix}
The evolution equation for the logarithm of the characteristic function $\Phi(k)$ is given in the main text Eq.~\eqref{Eq:Charac_Func}:
\begin{equation}
    \partial_t\Phi = ik[H - \partial_k^2\Phi - (\partial_k\Phi)^2] - |k|\Delta - (1-c)Dk^2,
\end{equation}
where $H = J r(t) + I_0 + \bar{\eta} + \sqrt{c}\xi_c(t)$.
Following Ref.~\cite{goldobin2021reduction}, $\Phi(k)$ can be expanded in the polynomial form
\begin{equation}
    \Phi = \sum_{n=1}^{\infty} \frac{q_n |k|^n + i p_n |k|^{n-1} k}{n},
    \label{Eq:Charac_Poly}
\end{equation}
with $q_1 = \pi r$ and $p_1 = -v$. 
The expansion in powers of $|k|$ (instead of $k$) reflects that the Lorentzian, serving as the reference state, lacks finite moments and cumulants. The pseudo-cumulant coefficients $W_n$ then describe deviations from this Lorentzian baseline, which are introduced as~\cite{goldobin2021reduction}
\begin{equation}
    W_1 \equiv \pi r - i v, \qquad W_n \equiv q_n + i p_n.
    \label{Eq:Pseudo_Cumu}
\end{equation}
Substituting the polynomial form~\eqref{Eq:Charac_Poly} into the evolution equation for $\Phi$ and using the definition of pseudo-cumulants~\eqref{Eq:Pseudo_Cumu}, one obtains a closed hierarchy for $W_m$. 
Specifically, differentiating $\Phi$ with respect to time and matching coefficients of equal powers of $|k|$ yields
\begin{align}
    \dot{W}_m &= (\Delta - i H)\delta_{1m} + 2(1-c)D\,\delta_{2m} \nonumber \\
    &\quad + i m \!\left[ -m W_{m+1} + \sum_{n=1}^{m} W_n W_{m+1-n} \right],
    \label{Eq:Evol_Cumu}
\end{align}
where $H$ is as defined above. 
Because higher-order pseudo-cumulants decay rapidly and have subleading effects on the macroscopic dynamics~\cite{goldobin2021reduction}, we retain only the first two, leading to
\begin{subequations}
\begin{align}
    \dot{r} &= (\Delta + p_2)/\pi + 2 r v, \label{Eq:rvqp_r} \\
    \dot{v} &= v^2+I_0 + \bar{\eta} + J r - \pi^2 r^2 + q_2 + \sqrt{c}\xi_c(t), \label{Eq:rvqp_v} \\
    \dot{q}_2 &= 2(1-c)D + 4 (p_3 + q_2 v - \pi p_2 r), \label{Eq:rvqp_q} \\
    \dot{p}_2 &= 4 (-q_3 + \pi q_2 r + p_2 v). \label{Eq:rvqp_p}
\end{align}
\end{subequations}
Because the second pseudo-cumulants $q_2$ and $p_2$ typically relax more quickly than the variables $r$ and $v$, as shown in Fig.~\ref{Fig:Trunc_ansatz} in the main text, i.e. $q_2$ and $p_2$ behave as fast modes slaved to the slow macroscopic variables $r$ and $v$. It is therefore reasonable to further assume a quasi-stationary closure $\dot{q}_2 = \dot{p}_2 = 0$.
Neglecting the third cumulant ($q_3 = p_3 = 0$) for Eqs.~\eqref{Eq:rvqp_q} and \eqref{Eq:rvqp_p}, we obtain
\begin{equation}
    p_2 = \frac{\pi r (1-c)D}{2(v^2 + \pi^2 r^2)}, 
    \qquad 
    q_2 = -\frac{v (1-c)D}{2(v^2 + \pi^2 r^2)}.
\end{equation}
Substituting these expressions into Eqs.~\eqref{Eq:rvqp_r} and~\eqref{Eq:rvqp_v}, we arrive at the macroscopic equation~\eqref{Eq:Langevin_r_v} presented in the main text.

\normalem
\bibliography{references}

\end{document}